\documentclass[conference]{IEEEtran}

\usepackage{cite}
\usepackage{amsmath,amssymb,amsfonts}
\usepackage{algorithmic}
\usepackage{graphicx}
\usepackage{textcomp}
\usepackage{comment}
\usepackage{xcolor}
\usepackage{color}

\def\BibTeX{{\rm B\kern-.05em{\sc i\kern-.025em b}\kern-.08em T\kern-.1667em\lower.7ex\hbox{E}\kern-.125emX}}
\usepackage{siunitx}
\usepackage{amsmath}
\usepackage{hyperref}

\begin{document}

\title{Design of a Canted-cosine-theta orbit corrector for the High Luminosity LHC}

\author{K. Pepitone, G. Kirby, R. Ruber, A. Ahl, M. Canale, I. Dugic, L. Gentini, \\ M. Johansson, G. Karlsson, J. Kovacikova, J. Lindström, A. Olsson, M. Olvegård}

\maketitle

\begin{abstract}
The High Luminosity LHC requires dipole orbit correctors grouped in double aperture magnet assemblies. They provide a field of 3.1 T at 100 A in an aperture of 70 mm. The current standard design is a classical cosine-theta layout made with ribbon cable. However, the electric insulation of the ribbon cable is not radiation-resistant enough to withstand the radiation load expected in the coming years of LHC operation. A new design, based on a radiation-resistant cable with polyimide insulator, that can replace the existing orbit correctors at their end-of-life, is needed. The challenge is to design a magnet that fits directly into the existing positions and that can operate with the same busbars, passive quench protection, and power supplies as existing magnets. We propose a self-protected canted-cosine-theta (CCT) design. We take the opportunity to explore new concepts for the CCT design to produce a cost-effective and high-quality design with a more sustainable use of resources. The new orbit corrector design meets high requirements on the field quality while keeping within the same mechanical volume and maximum excitation current. 

A collaboration of Swedish universities and Swedish industry has been formed for the development and production of a prototype magnet following a concurrent engineering (CE) methodology to reduce the time needed to produce a functional CCT magnet. The magnet has a 1\,m long CCT dipole layout consisting of two coils. The superconductor is a commercially available 0.33\,mm wire with polyimide insulation in a 6-around-1 cable. The channels in the coil formers, that determine the CCT layout, allow for $2\times5$ cable layers. A total of 70 windings makes that the coil current can be kept below 100 A. We will present the detailed design and preliminary quench simulations.
\end{abstract}


\date{\today}

\section{Introduction}

The existing orbit corrector magnets in the LHC, called MCBC and MCBY \cite{Bruning:782076}, are slated for replacement because of the radiation damage caused by the proton beams in the accelerator to the ribbon cables \cite{Louzguiti}. The CCT (Canted-Cosine-Theta) architecture was chosen for the new designs \cite{MEYER1970339}. Originally developed in the 1960s, the CCT design is based on the superposition of two solenoids tilted in opposite directions with respect to the bore axis. The choice of a radiation resistant polyimide insulated cable will give the correctors a longer life. CCT magnets offer advantages such as simplified manufacturing process and require less drawings than standard magnets. 

In this paper, we first present, Sec.~\ref{Magnet design}, the magnet parameters of the currently installed magnets, which will impose the conditions for the new design. Sec.~\ref{Simulations} focuses on the field simulations using COMSOL \cite{Comsol} and the discharge behavior using ProteCCT \cite{ProteCCT, Mentik}. Finally, in Sec.~\ref{CAD}, we discuss the engineering drawings and the mechanical assembly of this magnet.

\section{Magnet design} \label{Magnet design}
The characteristics of the original MCBC and MCBY magnets are shown in Table~\ref{tab:table1}. We intend to provide a one-to-one replacement magnet. The essential parameters to achieve this goal are highlighted in the same table. Note that the cold inner coil diameter is 70\,mm, so the warm diameter is 70.3\,mm using the aluminum 4.0\,mm/m contraction.

The power supplies used for the existing magnets can provide a maximum current of 120\,A \cite{Powersupplies}. Based on the first calculations we would need a current of approximately 600\,A and 10 conductors in each channel to obtain an integrated field equal to 2.81\,Tm for a magnetic length equal to 0.9\,m. To comply with the limitations posed by the power supply we increased the number of conductors by a factor 7. After discussion with superconducting (SC) cable manufacturers, we found the possibility to make cables composed of wires having a 6-around-1 pattern. The 10 cables of 7 wires will be connected in series and through this the current in each individual wire is reduced from 600 A to around 85 A. However, this new design is very demanding in terms of cabling since a total of 70 joints must be realized on each coil.

\begin{table*}[htbp]
\caption{MCBC and MCBY parameters.}
\begin{center}
\begin{tabular}{lcc}
\textbf{Parameters} & \textbf{MCBC} & \textbf{MCBY}\\
\hline
Coil inner diameter & 56.0\,mm & \textbf{70.0\,mm}\\
Magnetic length & 0.904\,m & 0.899\,m\\
Nominal field (at 1.9\,K / 4.2\,K) & \textbf{3.11\,T / 2.33\,T} & 3.00\,T / 2.50\,T\\
Integrated field (at 1.9\,K / 4.2\,K) & \textbf{2.81\,Tm / 2.10\,Tm} & 2.70\,Tm / 2.25\,Tm\\
Nominal current (at 1.9\,K / 4.2\,K) & \textbf{100\,A / 74\,A} & 88\,A / 72\,A\\
Short sample current (at 1.9\,K / 4.2\,K) & 172\,A / 127\,A & 162\,A / 120\,A\\
Stored energy (at 1.9\,K) & 14.2\,kJ & 13.6\,kJ\\
Self-inductance & 2.84\,H & 5.27\,H\\
DC resistance (RT) & 375\,\si{\ohm} & 501\,\si{\ohm}\\
Overall length & \textbf{1.1\,m} & \textbf{1.1\,m}\\
\end{tabular}
\label{tab:table1}
\end{center}
\end{table*}

\section{Simulations}\label{Simulations}

The starting point of the simulations is based on the work already realized at CERN \cite{kirby}. The initial work was realized using MATLAB and the toolbox \textit{Magnetic field of modulated double helical coils} \cite{queval}. This toolbox solves, in a few seconds, the Biot and Savart equation to compute the magnetic field. Based on the initial results, more detailed simulations of the field and of the discharge behavior were performed.

\subsection{Field simulations}\label{Field simulations}

The exact geometry of the magnet is defined in the COMSOL program including the number of wires in the cable, dimensions of the cable, total current and type of material for the formers and the yoke. The formers and the support are in aluminum; the superconductors are in Niobium-Titanium (NbTi) and the yoke in iron. All the parameters used for the simulations are given in Table~\ref{tab:table2}.

\begin{figure}[htbp]
\centering
\includegraphics[width = 8cm]{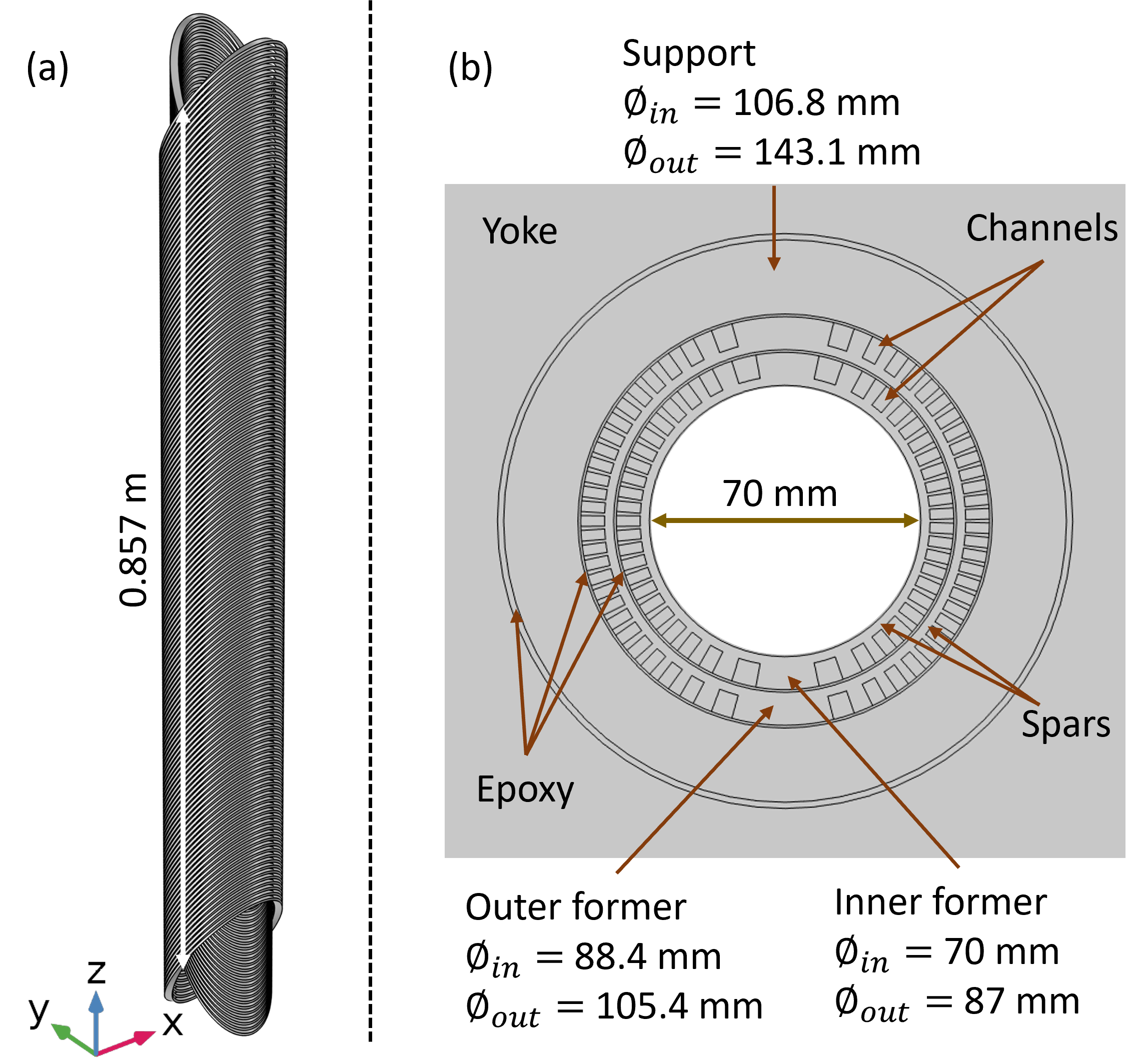}
\caption{3D view (a) and 2D view (b) of the model defined in COMSOL.}
\label{fig:COMSOL}
\end{figure}

\begin{figure}[htbp]
\centering
\includegraphics[width = 8cm]{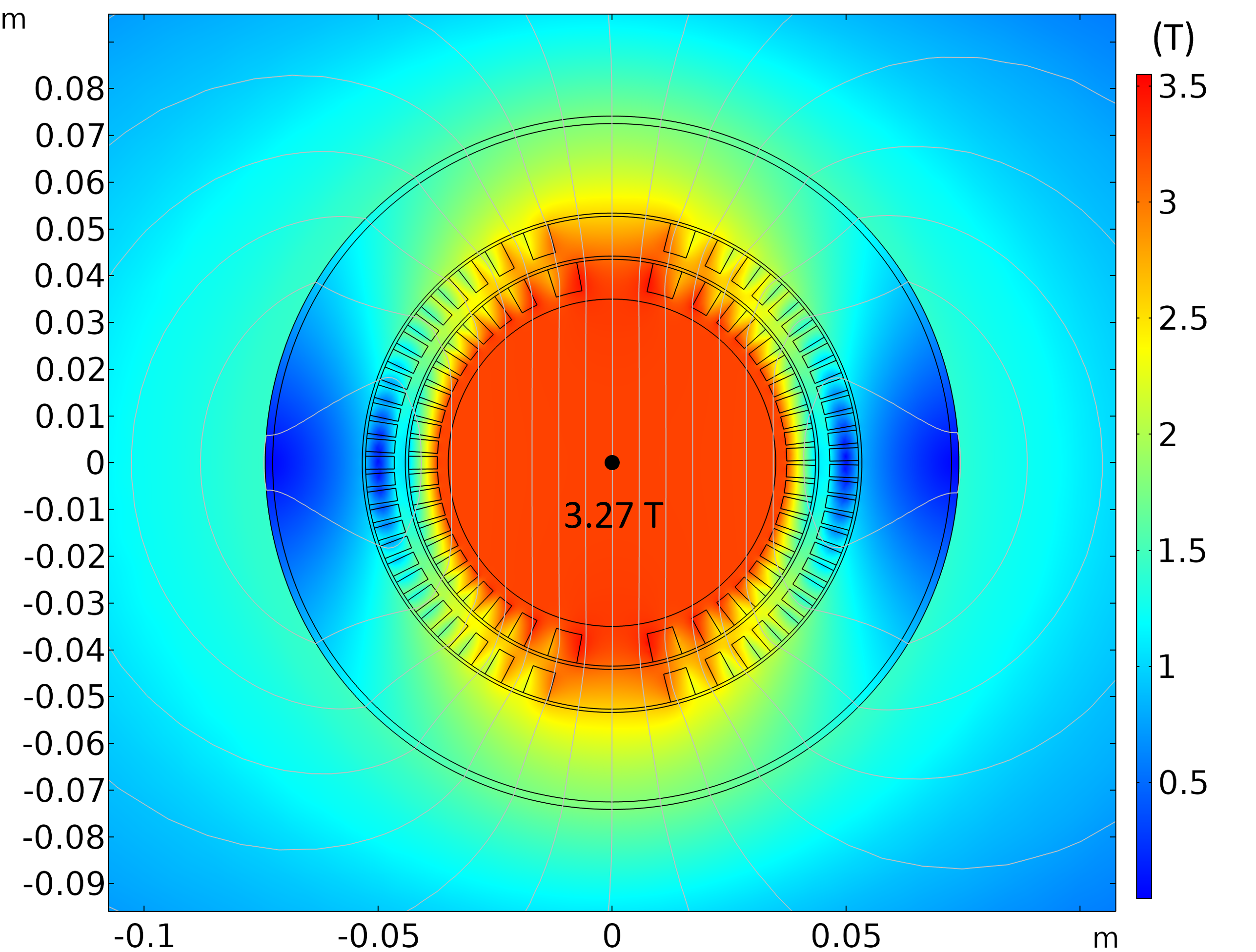}
\caption{2D view of the magnetic field simulated with COMSOL for a current of ${I}$ = 85\,A.}
\label{fig:Field}
\end{figure}

\begin{figure}[htbp]
\centering
\includegraphics[width = 8cm]{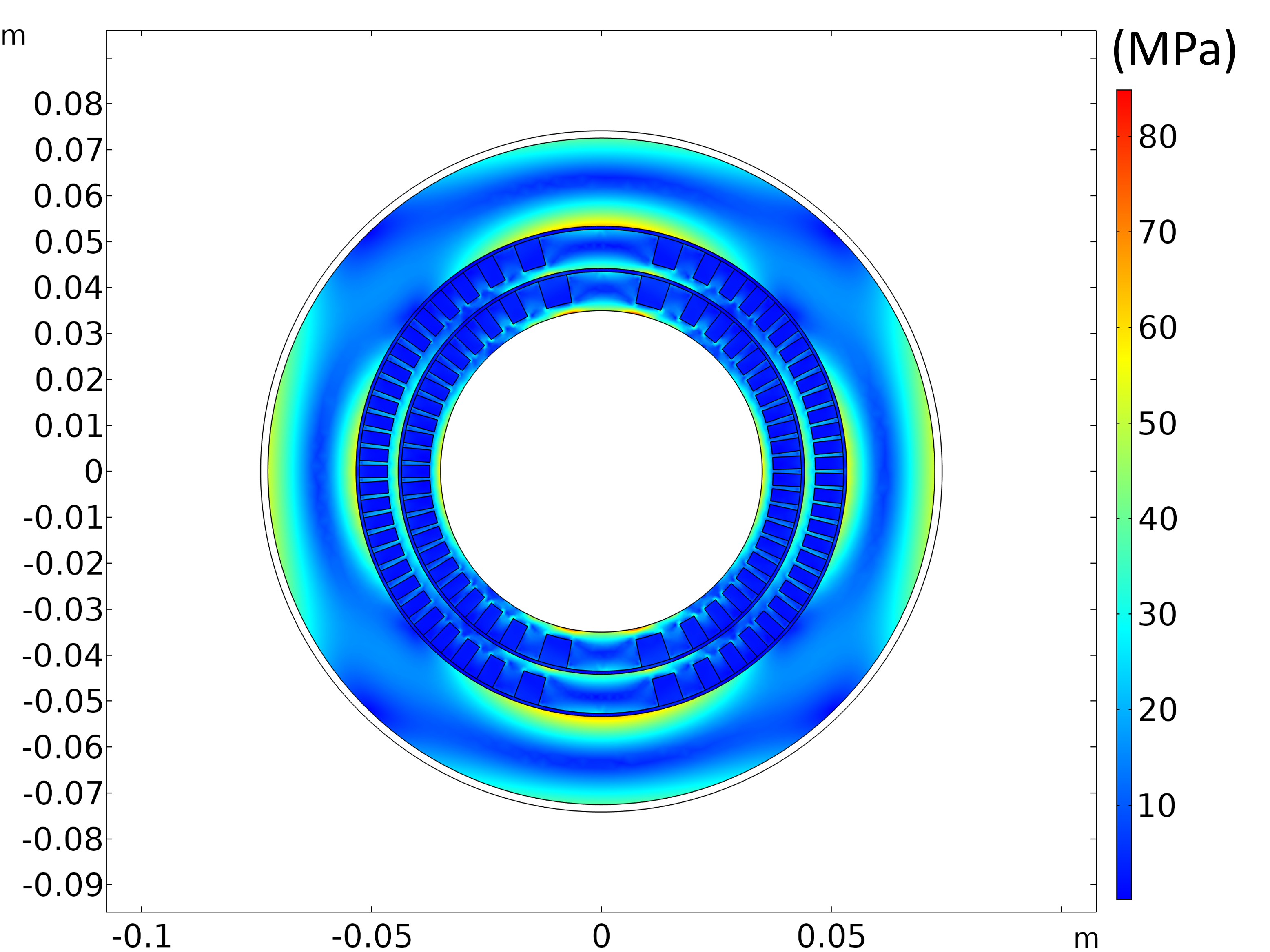}
\caption{2D view of the mechanical stress simulated with COMSOL for a current of ${I}$ = 85\,A.}
\label{fig:Stress}
\end{figure}

The current design is shown in Fig.~\ref{fig:COMSOL} and the simulation results are presented in Fig.~\ref{fig:Field} and Fig. ~\ref{fig:Stress}. 

\begin{table}[htbp]
\begin{center}
\caption{Parameters used in the simulations.}
\begin{tabular}{lc}
\textbf{Parameters} & \textbf{Values}\\
\hline
Coil inner diameter & \textbf{70.0\,mm}\\
Overall length & \textbf{1.1\,m}\\
Integrated field & \textbf{2.81\,Tm}\\
Magnetic length & 0.857\,m\\
Nominal current (per wire) & 85\,A\\
SC cable configuration & 6-around-1\\
SC cable diameter & 1.08\,mm \\
SC cable length & 1310\,m \\
SC wire (with / without insulation) & NbTi 0.36\,mm / 0.33\,mm\\
Cu to SC ratio & 1.3 $\pm$ 0.1 : 1\\
Critical current (at 4.2\,K) & 137\,A\\
Inner coil inner diameter & 75.7\,mm\\
Inner coil outer diameter & 87.0\,mm\\
Outer coil inner diameter & 94.1\,mm\\
Outer coil outer diameter & 105.4\,mm\\
Pitch & 5.8\,mm\\
Number of turns & 147\\
Tilt angle & 30$^{\circ}$\\
Thickness of the formers spars & 2.85\,mm\\
Wall thickness (min) & 0.3\,mm\\
Channel design & 2 x 5 cables\\
Channel dimensions & $2.31\times5.65$\,mm$^{2}$\\
\end{tabular}
\label{tab:table2}
\end{center}
\end{table}

For a total current of 85\,A per wire, a 6-around-1 cable and 10 cables in each channel, a central field of 3.27\,T is obtained. Considering that the magnetic length is 0.857\,m for an overall length below 1.1\,m, we obtain an integrated field of 2.8\,Tm. The final integrated field value, which must meet the specifications, can be reached by adjusting the current during the magnetic measurements of the first prototype at warm and at cold.

The stress analysis extracted from the simulations show that the mechanical stress on the formers is very low compared to what aluminum can withstand. We intend to use aluminum of the type AL6082 with a T6 or T4 heat treatment which has a stress yield of about 400\,MPa at room temperature and 600\,MPa when cold. Our simulations show that the maximum stress should be less than 70\,MPa.

\subsection{Quench simulations}\label{Quench simulations}

\begin{figure}[htbp]
\centering
\includegraphics[width = 8.7cm]{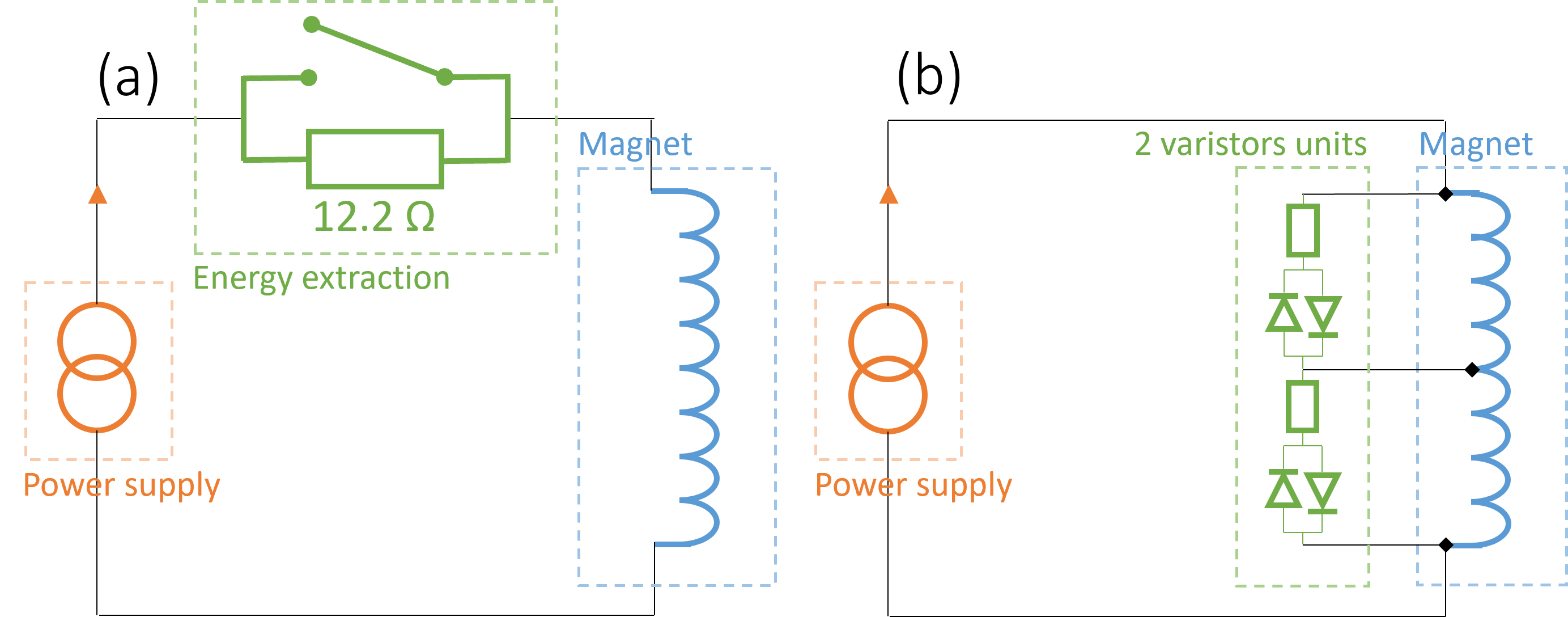}
\caption{Standard energy extraction circuit (a), and example passive extraction circuit using 2 varistors (b).}
\label{fig:Extraction}
\end{figure}

The MCBC/MCBY magnets do not have an energy extraction system which means that our prototype must be self-protected. A solution chosen to passively protect is to use varistors \cite{Varistors}. Varistors would have to be installed along the magnet for every 2500\,m of cable. If the voltage increases above the varistor threshold, the diode becomes conductive, and the energy of the magnet is extracted through the varistor dump resistor and dissipated in the helium bath. This option for quench protection is illustrated Fig.~\ref{fig:Extraction}~-~b and details on the varistors including a comparison with usual resistors can be found here \cite{Varistors}.

An alternative idea of using the central wire of the 6-around-1 cable as a quench heater or quench detector is also being considered. Tests on the first prototype will give an idea of the exact current needed to reach the 2.81\,Tm. Based on the results of these tests, we will decide if the central wire of each cable should remain connected in series with the others or if it can be used for one of the above mentioned applications.

Even if the magnet has to be self-protected, prototype will be tested with a standard energy extraction system (see Fig.~\ref{fig:Extraction}~-~a). Simulations have been performed to identify the resistor size and to check hot spot temperature.

The simulations were performed using the ProteCCT~\cite{ProteCCT, Mentik} software. ProteCCT calculates the eddy currents in the formers and in the support. All heat transfers are considered by the three-dimensional model, as well as heat exchanges with the helium bath and with the liquid helium present between the formers. The different couplings between the superconductors, the formers and the quench propagation are also considered. A description of the software and the physics implementation can be found here: \cite{ProtecctManual}.

The geometry, in terms of dimensions, number of conductors, excitation current and all the other parameters cited in Table~\ref{tab:table2} were fully defined in the toolbox. One of the problems of the implementation in ProteCCT is the geometry of the 6-around-1 cable. The heat transfer between the aluminum formers and the wires is not equal for the six twisted outer wires and the center wire. We decided to simulate a cable which is comparable to the 6-around-1 cable. This means that this cable has the same amount of superconductor, copper and insulation material as the 6-around-1 cable. The simulation therefore only involves 10 conductors and not 70 wires. With this approximation, we consider that the heat transfer is always the same. 

\begin{figure}[htbp]
\centering
\includegraphics[width = 8cm]{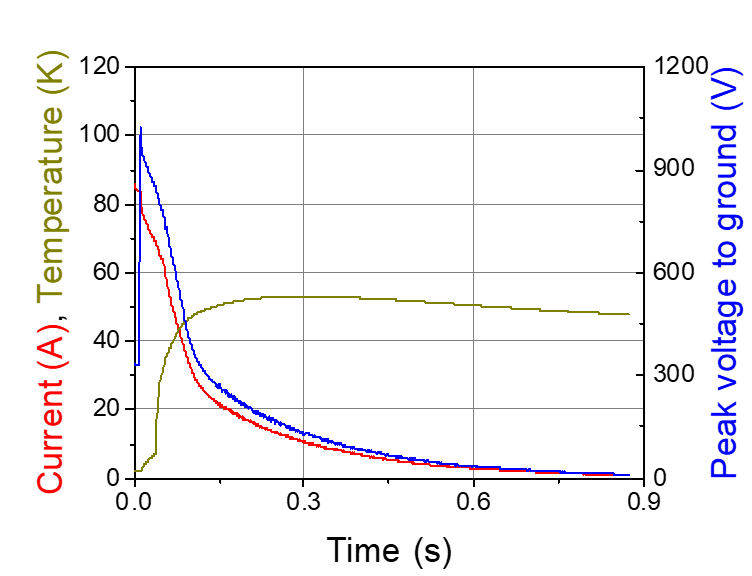}
\caption{Quench simulation for ${I}$ = 85\,A and a dump resistor of 12.2\,\si{\ohm}.}
\label{fig:Protecct}
\end{figure}

The results of the simulation presented in Fig.~\ref{fig:Protecct} are post-processed and correspond to what a single wire would see for a load of 12.2\,\si{\ohm}. With this discharge resistance, the maximum voltage to ground is just below 1000\,V, which is the limit for the equipment connected to the magnet (acquisition system and power supply).

\begin{figure}[htbp]
\centering
\includegraphics[width = 8cm]{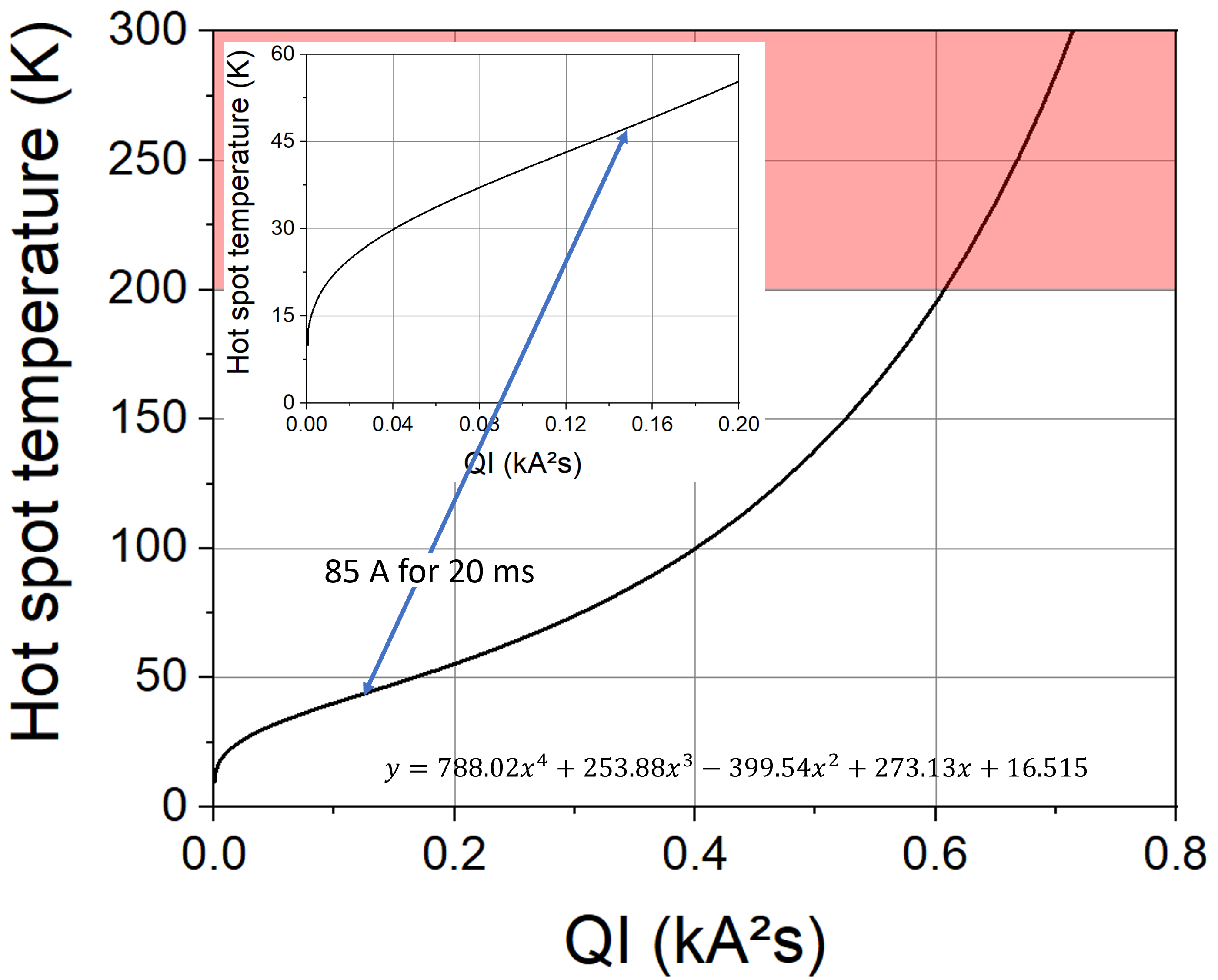}
\caption{Hot spot temperature in the conductor as a function of the current and the time.}
\label{fig:Hotspot}
\end{figure}

ProteCCT calculates the hot spot temperature at the time of discharge and not at the time of the initial quench. The detection voltage and the validation time will be set so that the discharge occurs at most 20\,ms after the start of the quench. Fig.~\ref{fig:Hotspot} shows the hot spot temperature in the conductor as function of a given current for a given time, extracted from a simple thermal model. If we assume a constant current of 85\,A during 20\,ms, we obtain a hot spot temperature increase of 47\,K. If we add this value to the hot spot temperature calculated by ProteCCT, shown in Fig.~\ref{fig:Protecct}, we obtain a total temperature of about 100\,K, which is below 200\,K, the empirical target hot spot temperature for this type of magnet.

\section{CAD model}\label{CAD}
\begin{figure}[htbp]
\centering
\includegraphics[width = 8cm]{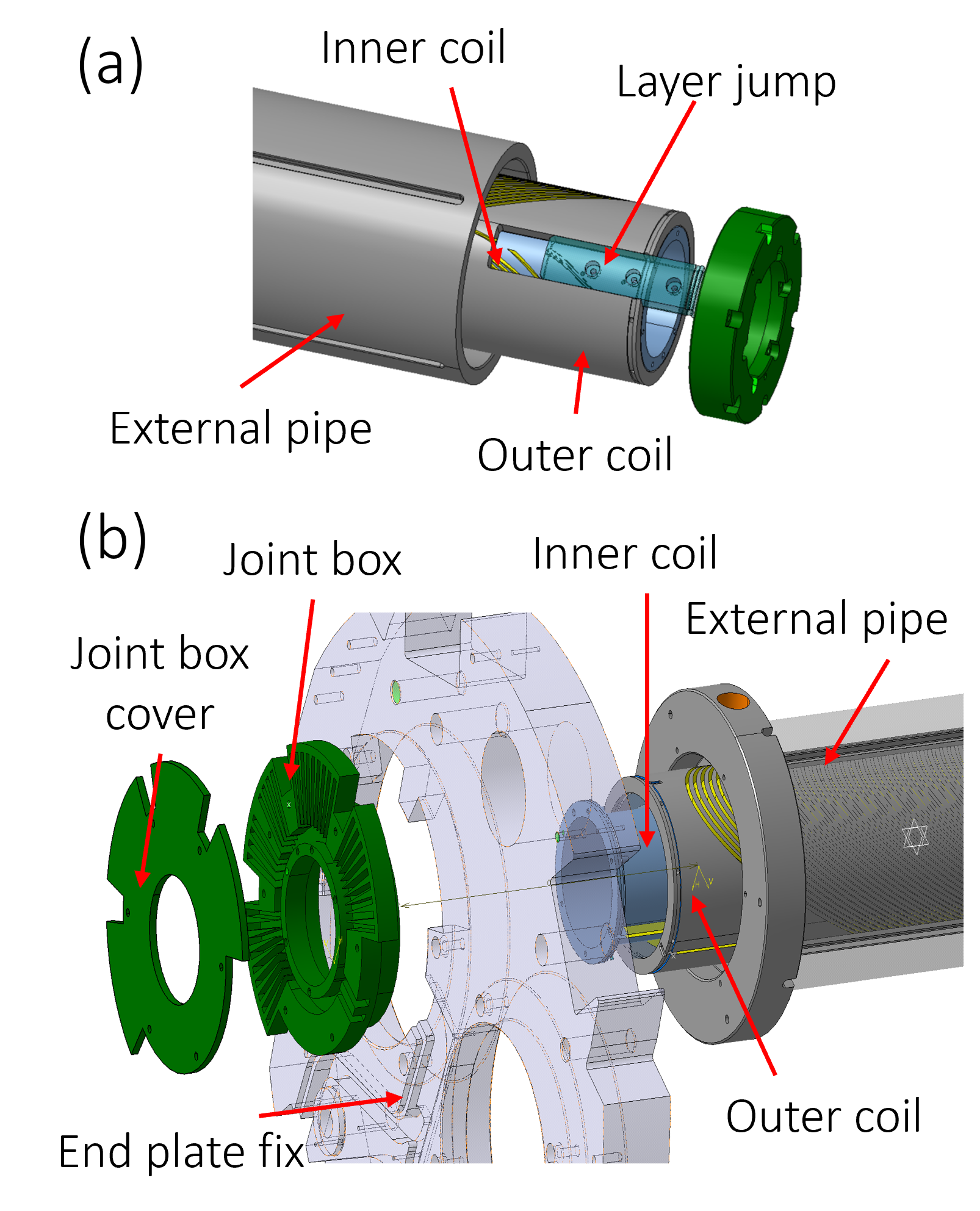}
\caption{3D view of the joint box (a) the layer jump (b).}
\label{fig:CAD}
\end{figure}

CAD (Computer-aided design) design is ongoing and include a number of new challenges compared to existing CCT magnets \cite{kirby}. These challenges are in three main areas: (a) the layer jump, where the 70 wires (10 cables) are brought from the inner former to the outer former (b) joint boxes, where all 70 individual wires are connected in series, and (c) the quench protection as described in Sec.~\ref{Quench simulations}. Based on the first experience of Scanditronix on assembling a CCT magnet there is need for improvement of the layer jump, so that the cables and their location in the channel are properly visible during assembly. We will use 3D printed parts of the CAD model to develop a good design for this section. See Fig.~\ref{fig:CAD}~-~a for an example of the layer jump.

A multi-layer joint box will be designed and installed on the side of the magnet. It will consist of three layers of G10 (high-pressure fiberglass laminate) similar to what is shown in Fig.~\ref{fig:CAD}~-~b. Two of these are used for the 70 splices and the last one for the instrumentation and voltage taps. These elements will not be impregnated and will be easily accessible in order to facilitate modifications or fixes, for example a short circuit, on the splices and connections.

\section{Conclusion}\label{Conclusion}

The CCT magnet being designed and manufactured by two Swedish universities and Swedish industry is now fully simulated. The model will be ready for manufacturing in the next few weeks and will be tested at FREIA laboratory at Uppsala University.
This magnet is based on the LHC MCBY and MCBC specifications and has three major improvements compared to other CCT magnets: improved layer jump, easily accessible joint boxes, and quench protection.

\section*{Acknowledgement}\label{Acknowledge}
This project has received funding from the European Regional Development Fund and Region Kronoberg.

The authors would like to express their gratitude to Matthias Mentink, Martin Novak, Emmanuele Ravaioli, and Mariusz Wozniak for their valuable help on the quench protection studies and COMSOL simulations.

\clearpage 
\bibliography{refs}
\bibliographystyle{unsrt}

\end{document}